\documentclass[preprint,11pt]{elsarticle}

\usepackage[a4paper, total={7.5in, 10.7in}]{geometry}
\usepackage{amssymb}
\usepackage{lscape}
\usepackage{graphicx}
\usepackage{tabularx}
\usepackage{float}
\usepackage{hyperref}
\usepackage{mathtools}

\usepackage{booktabs}

\newtheorem{theorem}{Theorem}
\newtheorem{example}{Example}

\begin{document}

\begin{frontmatter}



\title{Fractality in a delayed-prey-predator ecological system with Allee effects}

\author[inst1]{Sujay Goldar}
\ead{sujaygoldar@pinglacollege.ac.in}

\author[inst1]{Sk. Sarif Hassan\corref{cor1}}
\ead{sksarifhassan@pinglacollege.ac.in}
\cortext[cor1]{Corresponding author}

\affiliation[inst1]{organization={Department of Mathematics},
            addressline={Pingla Thana Mahavidyalaya}, 
            city={Maligram, Paschim Medinipur},
            postcode={721140}, 
            state={West Bengal},
            country={India}}

\begin{abstract}
A delayed, discrete-time, prey-predator model with Allee effects imposed on prey and predator populations is defined, and dynamics of the system is characterized computationally. The parametric conditions for local asymptotic stability of equilibria are investigated with several illustrating examples. In addition, several periodic solutions were achieved. Furthermore, high degree of fractality of the prey-predator population trajectories is observed when the system is sufficiently delayed. This system has turned out to be a cooperative system, which is depicted using fractal dimension. 
\end{abstract}


\end{frontmatter}


\section{Introduction}
Many of the ecological dynamics models deal with a single population dispersing among patches \cite{ghosh2017prey, saez1999dynamics, dixon2000insect, xiao2001global}. Few of the models deal with competition and predator-prey interactions in patchy environments \cite{sarwardi2012analysis, wang2008positive, sih1985predation}. Analysis of such a model is to study the coexistence of populations and the stability (local and global) of equilibria \cite{meng2018dynamics, zhang2017dynamics, danca1997detailed, hassan2018computational, hassan2018dynamics}. Prey-predator interaction is one of basic interspecies relations for ecological models, food chain models, and biochemical network structure \cite{huang2018dynamical}. Notion of Allee effect was introduced by Warder Clyde Allee, an American ecologist, and later it was widely by several authors \cite{wang2011predator, drake2011allee, petrovskii2002allee, celik2009allee, zhou2005stability, wang2011analysis}. The Allee effect may arise from a number of sources such as difficulties in finding mates, reproductive facilitation, predation, and environment conditioning \cite{stephens1999allee, sun2016mathematical, scheuring1999allee}. Earlier, it was reported that delays have destabilizing effects on the prey-predator dynamics, which induce periodic solutions as well as chaotic solutions \cite{gakkhar2012complex, singh2019complex}. Delay arises due to several reasons such as gestation, hunting, maturation, and so on, in ecological systems \cite{macdonald2008biological}. For a long period, delay has been investigated in various prey predator models by plenty of researchers \cite{gakkhar2012complex, macdonald2008biological}. Allee effects is a phenomenon in biology characterized by a correlation between population size or density and the mean individual fitness of a population or species. Earlier, in the article authored by Q.Din presented a discrete model of prey-predator dynamics under the influence of Allee effects on prey population \cite{din2018controlling}. In this present study, we have modified and redefined the model with Allee effects imposed on both the populations. Note that, in addition, a delay $'f'$ is also introduced in the system.  \\

We have introduced the following model:

\begin{equation}
     x_{n+1}=x_{n-f}e^{\left(r(1-\frac{x_{n-f}}{k})\frac{x_{n-f}}{a+x_{n-f}}-by_{n-f}\right)}
\end{equation}
   
\begin{equation}
    y_{n+1}=y_{n-f}e^{\left(bx_{n-f}-\frac{d y_{n-f}}{c+y_{n-f}} \right)}
\end{equation}

\noindent
Here the parameters were defined as follows:\\
$r$: intrinsic growth rate of prey population,\\
$k$: carrying capacity of prey population,\\
$a$: allee constant imposed on prey,\\
$b$: predation rate,\\
$c$: allee constant imposed on predator,\\
and $d$: per capita mortality rate of predator.

\noindent
Note that all the parameters $r, k, a, b, c,$ and $d$ are non-negative real numbers. Here the variables $x$ and $y$ represent density of prey and predator respectively. \\

The article is organized as follows. In the next section we describe local asymptotic stability of equilibria, and periodic solutions. Furthermore, some special cases when either prey or predator Alee constant is zero, are presented in the last subsection of the Result.

\section{Results}

\subsection{Local Asymptotic Stability Analysis}
The equilibria of the system of Eqs.(1-2) are obtained by solving the following system:\\
$$\frac{r x \left(1-\frac{x}{k}\right)}{a+x}-b y=0$$
$$b x-\frac{d y}{c+y}=0$$
Hence the two equilibria are $O=(0,0)$ and $P=\left(\frac{-M+b k (b c+r)+d r}{2 b r}, \frac{d \left(b k (r-b c)-d r+M\right)-2 a b^3 c k}{2 b^2 k (a b+d)}\right)$. Note that $M=\sqrt{2 b^2 c k r (2 a b+b k+d)+b^4 c^2 k^2+r^2 (d-b k)^2}$. It is noted that the equilibrium $P$ of the system of Eqs.(1-2) exists uniquely if and only if $ab^2c<d r$.

Furthermore, Jacobian matrix
of system of Eqs.(1-2) at a point $(x,y)$ is given by $$\left(
\begin{array}{cc}
 \frac{\left(a^2 k+a x (k (r+2)-2 r x)+x^2 (k-r x)\right) e^{\frac{r x (k-x)}{k (a+x)}-b y}}{k (a+x)^2} & -b x e^{\frac{r x (k-x)}{k (a+x)}-b y} \\
 b y e^{b x-\frac{d y}{c+y}} & \frac{\left(c^2-c (d-2) y+y^2\right) e^{b x-\frac{d y}{c+y}}}{(c+y)^2} \\
\end{array}
\right)$$

\noindent
Therefore, the Jacobian at the trivial equilibrium $O$ is turned out to be $\left(
\begin{array}{cc}
 1 & 0 \\
 0 & 1 \\
\end{array}
\right)$ 
and hence the trivial equilibrium $O$ is an unstable source. 

\begin{example}

When $r=0.6135, k=0.5822, a=0.5407, b=0.8699, c=0.2648, d=0.3181$, and the delay $f=7$ then the origin $O$ is an unstable source, which forms a periodic cycle of period 4, as depicted in the Fig.\ref{fig:fg1}.

\begin{figure}
\begin{center}
\includegraphics[scale=0.55]{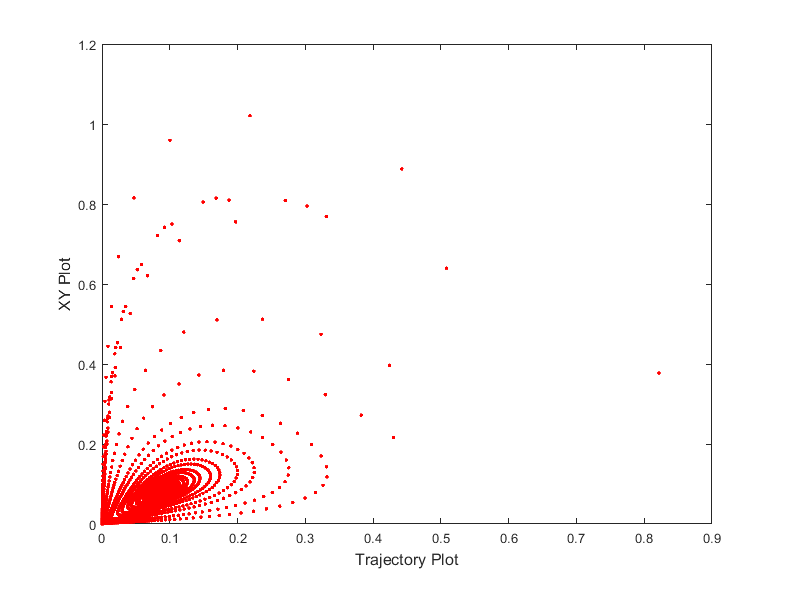}\\
\includegraphics[scale=0.45]{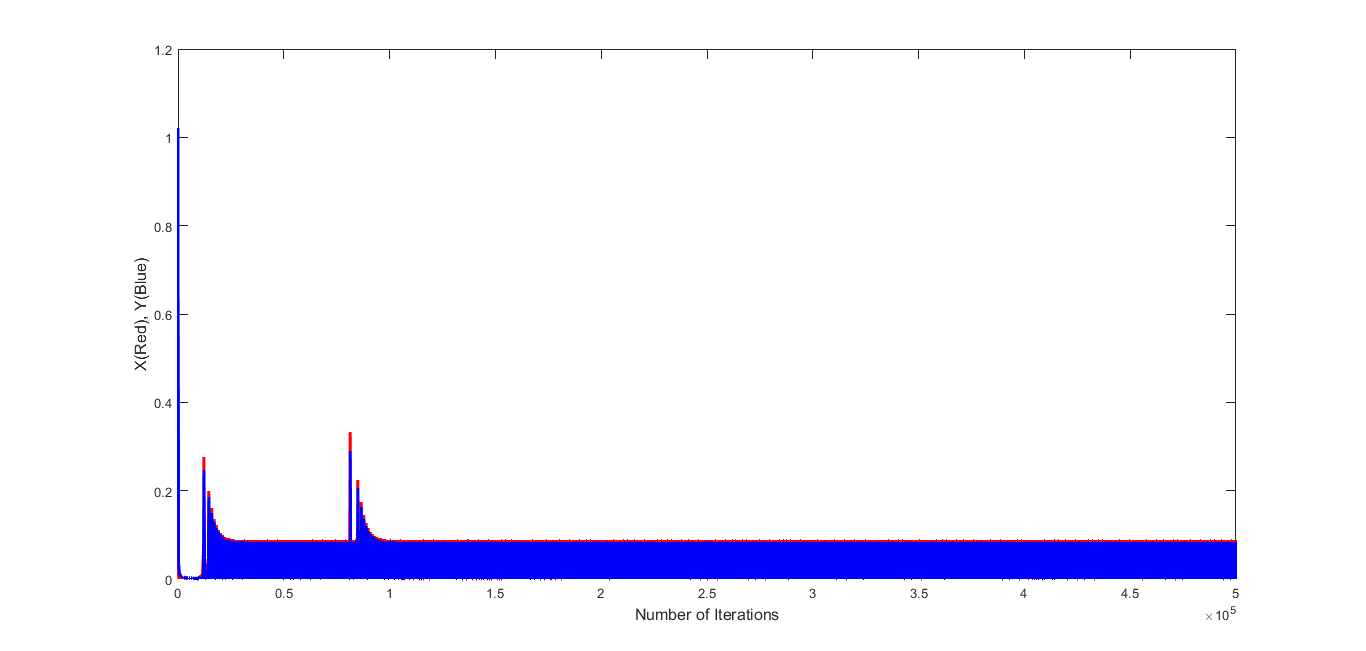}
\end{center}
\caption{Unstable source $O$ is leading to a periodic trajectory of period 4.}
\label{fig:fg1}
\end{figure}

\end{example}

\begin{example}

When $r=0.5755, k=0.5822, a=0.2751, b=0.2486, c=0.4516, d=0.2277$, and the delay $f=35$ then the origin $O$ is an unstable source which converges to the non-zero equilibrium $(0.3964, 0.3446)$, as depicted in the Fig.\ref{fig:fg2}.

\begin{figure}[H]
\begin{center}
\includegraphics[scale=0.43]{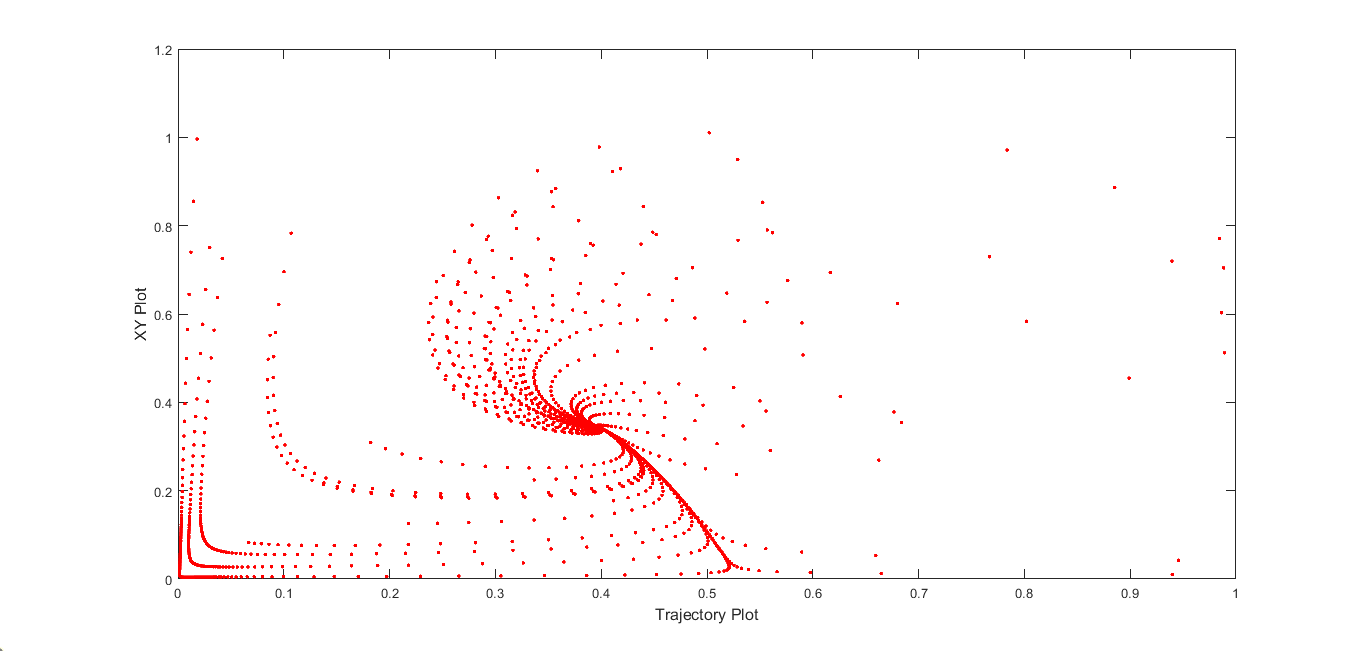}\\
\includegraphics[scale=0.43]{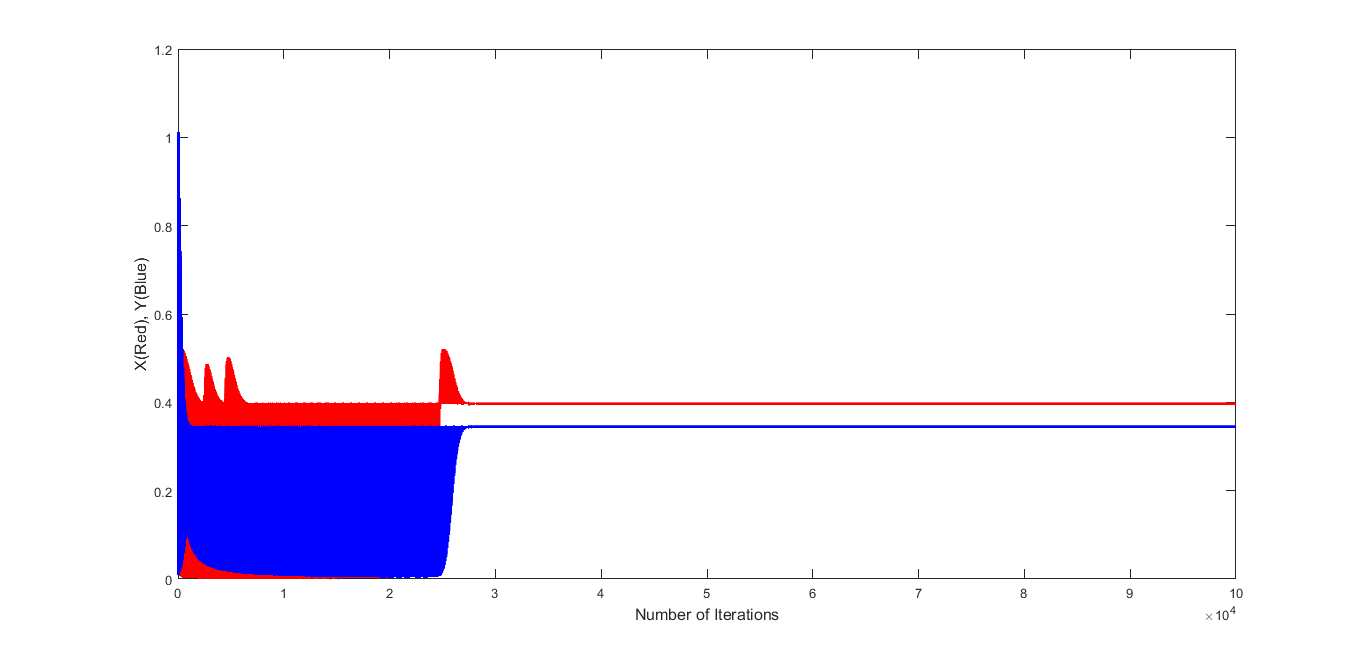}
\caption{Unstable source $O$ is leading to a convergent trajectory to the other equilibrium.}
\label{fig:fg2}
\end{center}
\end{figure}
\end{example}

\noindent
Now, about the non-trivial equilibrium $P=\left(\frac{-M+b k (b c+r)+d r}{2 b r}, \frac{d \left(b k (r-b c)-d r+M\right)-2 a b^3 c k}{2 b^2 k (a b+d)}\right)$, Jacobian matrix of system of Eqs.(1-2) becomes 
$\left(\begin{array}{cc}
 m_1 & m_2 \\
 m_3 & m_4 \\
\end{array}
\right)$. Note that, expressions for $m_1, m_2, m_3$ and $m_4$ are given in the \textit{Appendix-1}. The following theorem of the local asymptotic stability of the equilibrium $P$ is followed by the \textit{Result-1}.

\begin{theorem}
The equilibrium $P=\left(\frac{-M+b k (b c+r)+d r}{2 b r}, \frac{d \left(b k (r-b c)-d r+M\right)-2 a b^3 c k}{2 b^2 k (a b+d)}\right)$ is locally asymptotically stable if $$|m_1m_4-m_2m_3|+|m_1+m_4|<1$$
\end{theorem}

Following two examples with particular set of parameters illustrate the \textit{Theorem-1}.

\begin{example}
Let $r=0.4820, k=0.1206, a=0.5895, b=0.2262, c=0.3846, d=0.5830$ and the delay $f=27$, then the equilibrium $P=(0.1144, 0.0179)$ is locally asymptotically stable since the eigenvalues of the Jacobian \\ $\left(
\begin{array}{cc}
 -0.275376 & -0.0465808 \\
 0.0469666 & 0.807626 \\
\end{array}
\right)$ are $-0.2734,$ and $0.8056$. Note that, in this example, $|m_1m_4-m_2m_3|+|m_1+m_4|=0.312 < 1$, and consequently it agrees the condition for local asymptotic stability as stated in the \textit{Theorem-1}. \\

\noindent
The stable phase plot along with the trajectory plot of the prey ($x$) and predator ($y$) for the equilibrium $P=(0.1144, 0.0179)$ is given in Fig.\ref{fig:fg4}.

\begin{figure}[H]
\begin{center}
\includegraphics[scale=0.45]{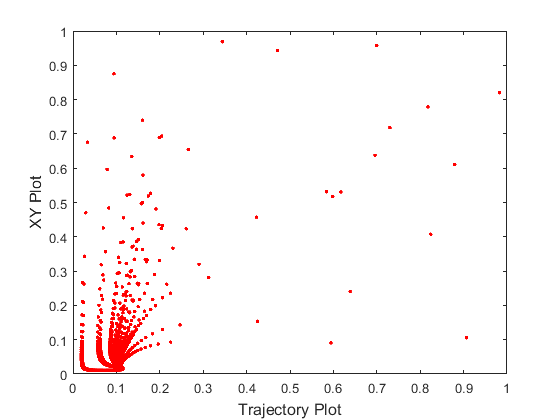}
\includegraphics[scale=0.45]{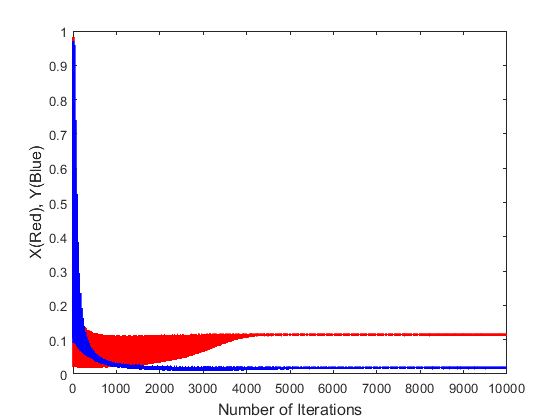}
\caption{Locally asymptotically stable phase plot with it trajectory plots of the equilibrium $P=(0.1144, 0.0179)$.}
\label{fig:fg4}
\end{center}
\end{figure}

\end{example}

\begin{example}
Let $r=0.9848, k=0.7157, a=0.8390, b=0.4333, c=0.4706, d=0.5607$ and the delay $f=14$, then the equilibrium $P=(0.4770, 0.2747)$ is unstable focus since the eigenvalues of the Jacobian $\left(
\begin{array}{cc}
 0.619298 & -0.240818 \\
 0.15894 & 0.88174 \\
\end{array}
\right)$ are both having positive real part $(0.7505 \pm 0.1451i)$. Note that, in this example, $|m_1m_4-m_2m_3|+|m_1+m_4|=2.0854 > 1$, and consequently it violets the condition for local asymptotic stability as stated in the \textit{Theorem-1}. \\

\noindent
The unstable phase plot for the equilibrium $P=(0.4770, 0.2747)$ is given in Fig.\ref{fig:fg3}.

\begin{figure}[H]
\begin{center}
\includegraphics[scale=0.75]{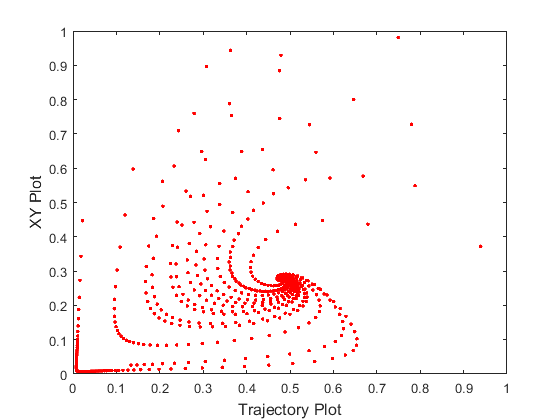}
\caption{Unstable phase plot of the equilibrium $P=(0.4770, 0.2747)$.}
\label{fig:fg3}
\end{center}
\end{figure}

\end{example}

\begin{example}
Let $r=0.9883, k=0.7668, a=0.3367, b=0.6642, c=0.2442, d=0.2955$ and the delay $f=71$, then the equilibrium $P=(0.28445, 0.4298)$ is unstable since the eigenvalues of the Jacobian $\left(
\begin{array}{cc}
 0.619298 & -0.240818 \\
 0.15894 & 0.88174 \\
\end{array}
\right)$ are $0.5991$ and $1.105$. Note that, in this example, $|m_1m_4-m_2m_3|+|m_1+m_4|=2.3661 > 1$, and consequently it violets the condition for local asymptotic stability as stated in the \textit{Theorem-1}. \\

\noindent
The unstable phase plot for the equilibrium $P=(0.28445, 0.4298)$ is given in Fig.\ref{fig:fg5}.

\begin{figure}[H]
\begin{center}
\includegraphics[scale=0.45]{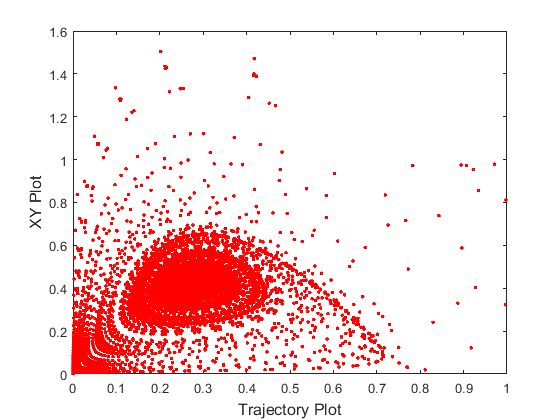}
\includegraphics[scale=0.45]{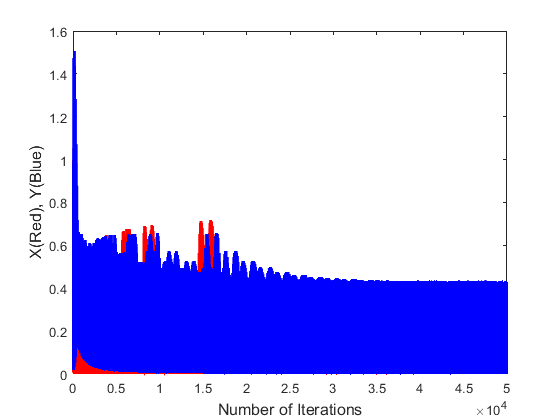}
\caption{Unstable phase plot of the equilibrium $P=(0.28445, 0.4298)$.}
\label{fig:fg5}
\end{center}
\end{figure}

\end{example}

\begin{example}
Let $r=4.9453, k=0.9482, a=0.7978, b=0.1330, c=0.2378, d=0.1277$ and the delay $f=55$, then the equilibrium $P=(0.853916, 1.91144)$ is unstable saddle since the eigenvalues of the Jacobian \\
$\left(
\begin{array}{cc}
 -1.17964 & -0.113571 \\
 0.254221 & 0.987434 \\
\end{array}
\right)$ are $-1.1662$ and $0.9740$. Note that, in this example, $|m_1m_4-m_2m_3|+|m_1+m_4|=1.3282 > 1$, and consequently it violets the condition for local asymptotic stability as stated in the \textit{Theorem-1}. \\

\noindent
The unstable saddle phase plot for the equilibrium $P=(0.853916, 1.91144)$ is given in Fig.\ref{fig:fg6}.

\begin{figure}[H]
\begin{center}
\includegraphics[scale=0.4]{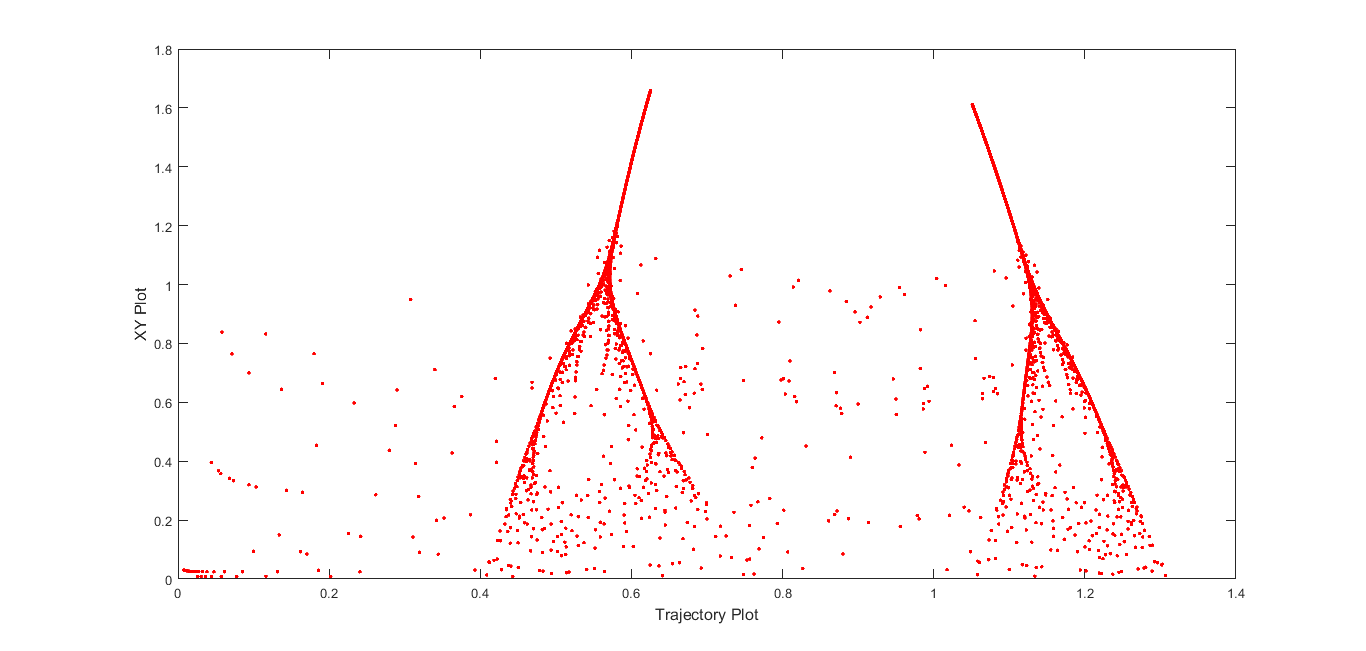}
\includegraphics[scale=0.4]{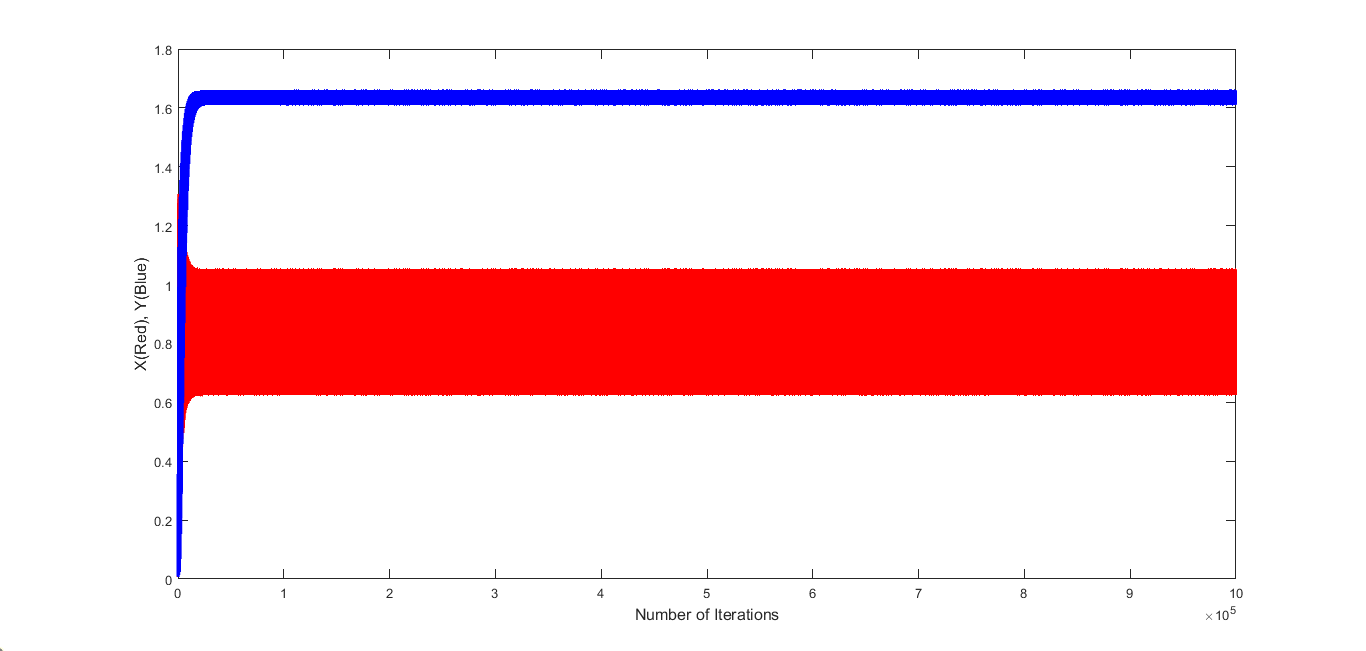}
\caption{Unstable (saddle) phase plot of the equilibrium $P=(0.853916, 1.91144)$.}
\label{fig:fg6}
\end{center}
\end{figure}

\end{example}

\subsection{Periodic Solutions}
Qualitatively, unstable equilibria emerge to periodic cycles. In this section, we shall explore possibility of periodic trajectories for different set of parameters.

\begin{table}[H]
\begin{center}
\caption{Parameters of the system Eqs.(1-2) and their associated trajectory characters.}
\begin{tabular}{@{}ccc@{}}
\toprule
\textbf{Parameters $(r, k, a, b, c, d, \textbf{f})$} & \textbf{Character}  & \textbf{Reference} \\ \midrule
$(0.6070, 1.2443, 0.5816, 0.8680, 0.1307, 0.3637, \textbf{5})$                 & Periodic                                              (High pericodicity) & Fig.\ref{fig:fg7} (Row-1) \\
$(0.5774, 0.6371, 0.9802, 0.1988, 0.2407, 0.0595, \textbf{83})$                 & Quasi-periodic                                               & Fig.\ref{fig:fg7} (Row-2)\\
$(0.8066, 0.8453, 0.2272, 0.9369, 0.4274, 0.4320 , \textbf{53})$                 & Quasi-periodic                                               & Fig.\ref{fig:fg7} (Row-3)\\ \bottomrule
\end{tabular}
\end{center}
\end{table}

\begin{figure}[H]
\begin{center}
\includegraphics[scale=0.55]{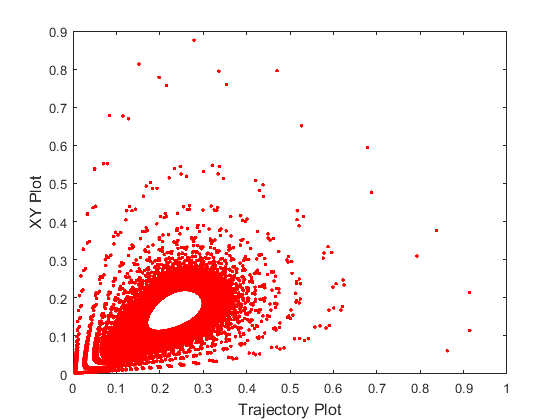}
\includegraphics[scale=0.55]{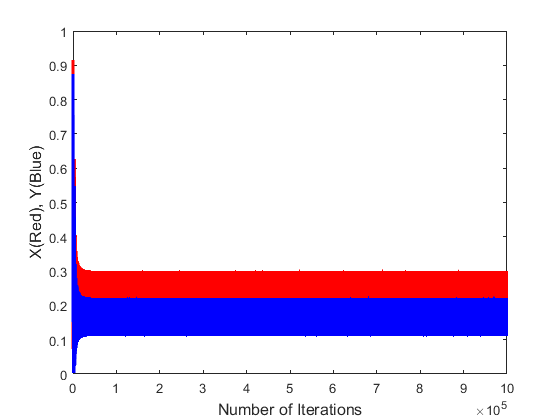}
\includegraphics[scale=0.55]{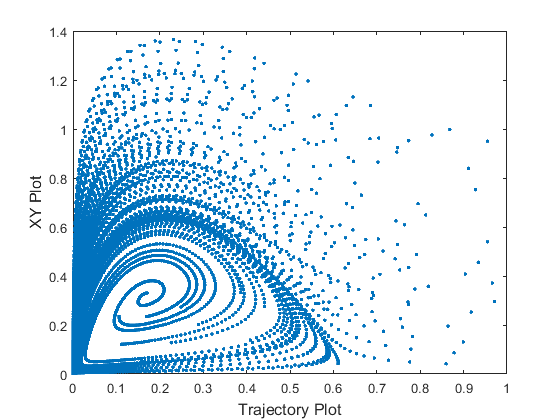}
\includegraphics[scale=0.55]{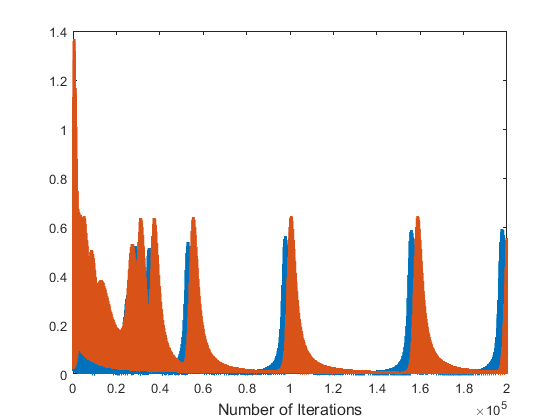}
\includegraphics[scale=0.55]{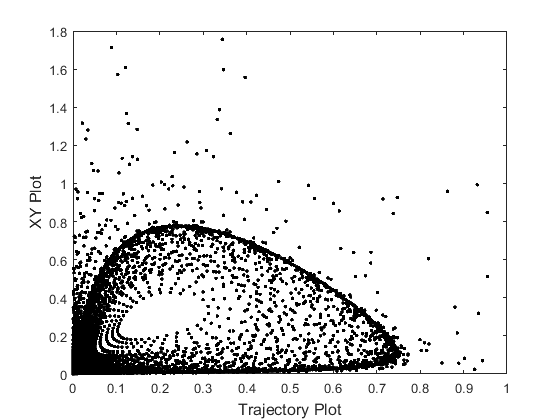}
\includegraphics[scale=0.55]{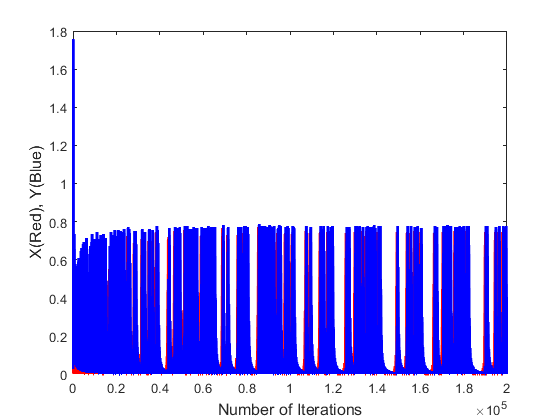}

\caption{Phase space and trajectory plot of quasi-periodic orbits}
\label{fig:fg7}
\end{center}
\end{figure}

In the three examples cited in Table 1 show periodic trajectories with high periodicities (Fig.\ref{fig:fg7})

\subsection{Special Cases}
In this section we shall explore the dynamical behaviour of the prey-predator system when the Allee constant is considered to be zero.

\subsection{When the Allee constant imposed on prey $a=0$}
when $a=0$, the positive equilibrium of the system of Eqs.(1-2) becomes $(s_1, s_2)$ where $$s_1=\frac{-\sqrt{b^4 c^2 k^2+2 b^2 c k r (b k+d)+r^2 (d-b k)^2}+b k (b c+r)+d r}{2 b r}$$ and $s_2=\frac{\sqrt{b^4 c^2 k^2+2 b^2 c k r (b k+d)+r^2 (d-b k)^2}+b k (r-b c)-d r}{2 b^2 k}$

Jacobian matrix of system of Eqs.(1-2) at a point $(x,y)$ is given by $$\left(
\begin{array}{cc}
 \frac{(k-r x) e^{-b y-\frac{r x}{k}+r}}{k} & -b x e^{-b y-\frac{r x}{k}+r} \\
 b y e^{b x-\frac{d y}{c+y}} & \frac{\left(c^2-c (d-2) y+y^2\right) e^{b x-\frac{d y}{c+y}}}{(c+y)^2} \\
\end{array}
\right)$$

\begin{theorem}
The equilibrium $(s_1, s_2)$ is locally asymptotically stable if $$|P|+|Q|<1$$ where P, the determinant of the Jacobian is given by $$\frac{e^{b (x-y)-\frac{d y}{c+y}-\frac{r x}{k}+r} \left(c^2 \left(b^2 k x y+k-r x\right)+c y \left(2 b^2 k x y+(d-2) (r x-k)\right)+y^2 \left(b^2 k x y+k-r x\right)\right)}{k (c+y)^2}$$ and Q, the trace of the Jacobian is given by $\frac{\left(c^2-c (d-2) y+y^2\right) e^{b x-\frac{d y}{c+y}}}{(c+y)^2}+\frac{(k-r x) e^{-b y-\frac{r x}{k}+r}}{k}$.
\end{theorem}

\noindent
\textit{Theorem-2} is illustrated through some examples presented as follows:\\

\begin{example}
Let $r=0.5, k=0.98, a=0, b=0.25, c=0.95, d=0.45$ and the delay f ranging from 1 to 50, then the equilibrium $P=(0.6904, 0.5911)$ is unstable (spiral source), since the eigenvalues of the Jacobian \\
$\left(
\begin{array}{cc}
 0.647766 & -0.172594 \\
 0.147766 & 0.893603 \\
\end{array}
\right)$ are $0.7707\pm0.1019i$ where real part are positive, and note that, in this example, $|P|+|Q|>1$. Therefore, it violets the condition for local asymptotic stability as stated in the \textit{Theorem-2}.

\noindent
The spiral source phase plot for the equilibrium $P=(0.6904, 0.5911)$ is given in Fig.\ref{fig:fg8}.

\begin{figure}[H]
\begin{center}
\includegraphics[scale=0.7]{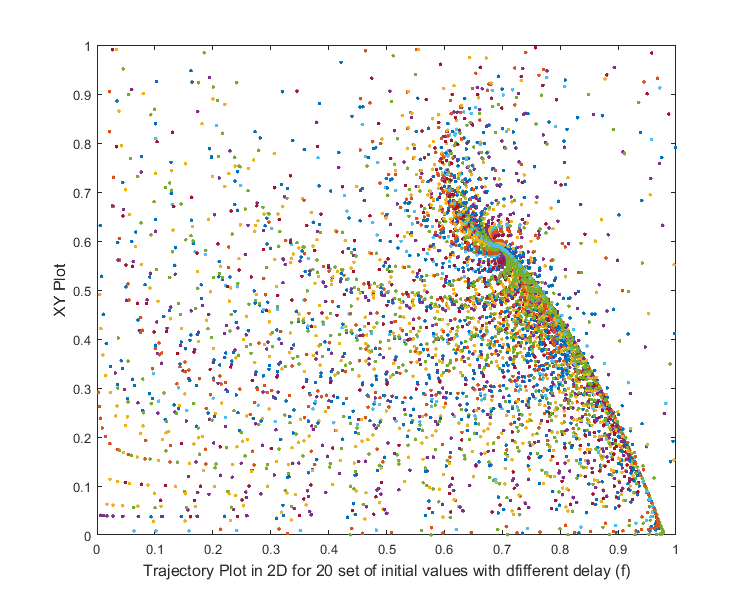}
\caption{spiral source phase plot of the equilibrium $P=(0.6904, 0.5911)$.}
\label{fig:fg8}
\end{center}
\end{figure}

\end{example}

\begin{example}
Let $r=0.5, k=0.98, a=0, b=0.25, c=0.95, d=0.45$ and the delay f ranging from 1 to 50, then the equilibrium $P=(0.2857, 0.2015)$ is locally asymptotically stable and parameters at the equilibrium satisfy the \textit{Theorem-2}. The trajectory phase plot for the equilibrium$P=(0.2857, 0.2015)$ is given in Fig.\ref{fig:fg9}.

\begin{figure}[H]
\begin{center}
\includegraphics[scale=0.8]{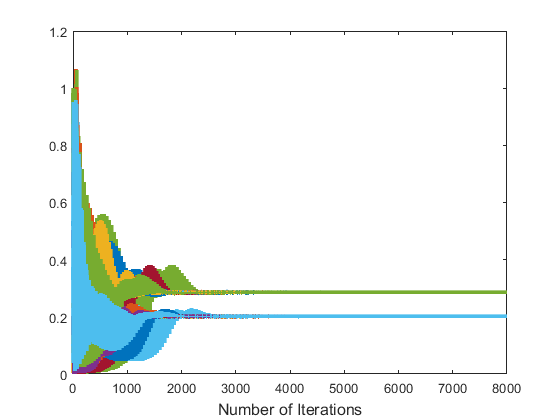}
\caption{Locally asymptotically stable phase plot of the equilibrium $P=(0.2857, 0.2015)$.}
\label{fig:fg9}
\end{center}
\end{figure}

\end{example}

\subsection{When the Allee constant imposed on predator $c=0$}
When $c=0$, the only positive equilibrium of the system of Eqs.(1-2) becomes $(\frac{d}{b}, \frac{d r (b k-d)}{b^2 k (a b+d)})$, provided $d<bk$.

Jacobian matrix of system of Eqs.(1-2) at the equilibrium $(\frac{d}{b}, \frac{d r (b k-d)}{b^2 k (a b+d)})$ becomes $$\left(
\begin{array}{cc}
 1-\frac{d r \left(a b (2 d-b k)+d^2\right)}{b k (a b+d)^2} & -d \\
 \frac{d r (b k-d)}{b k (a b+d)} & 1 \\
\end{array}
\right)$$

\begin{theorem}
The equilibrium $(s_1, s_2)$ is locally asymptotically stable if 

$$\left|1-\frac{d r \left(d \left(a b (d+2)+d^2+d\right)-b k \left(a b (d+1)+d^2\right)\right)}{b k (a b+d)^2}\right|+\left|2-\frac{d r \left(a b (2 d-b k)+d^2\right)}{b k (a b+d)^2}\right|<1$$ 
\end{theorem}

Here, one counter-example and an examples by illustrating \textit{Theorem-3} are given as follows:

\begin{example}
Let $r=0.9541, k=0.5428, a=0.5401, b=0.3111, c=0, d=0.0712$ and the delay f ranging from 1 to 50, then the equilibrium $P=(0.229, 0.528)$ is unstable (spiral source), since the eigenvalues of the Jacobian \\
$\left(
\begin{array}{cc}
 0.995623 & -0.0712 \\
 0.164235 & 1 \\
\end{array}
\right)$ are $0.997812\pm 0.108115 i$ where real part are positive, and note that, in this example, the parameters do not satisfy the condition stated in \textit{Theorem-3}. The unstable spiral source phase plot for the equilibrium $P=(0.229, 0.528)$ is given in Fig.\ref{fig:fg10}.

\begin{figure}[H]
\begin{center}
\includegraphics[scale=0.6]{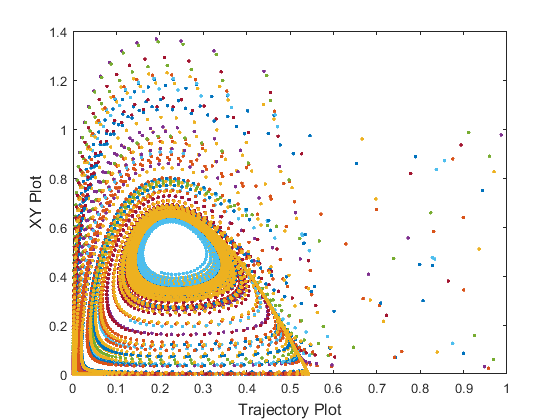}
\includegraphics[scale=0.25]{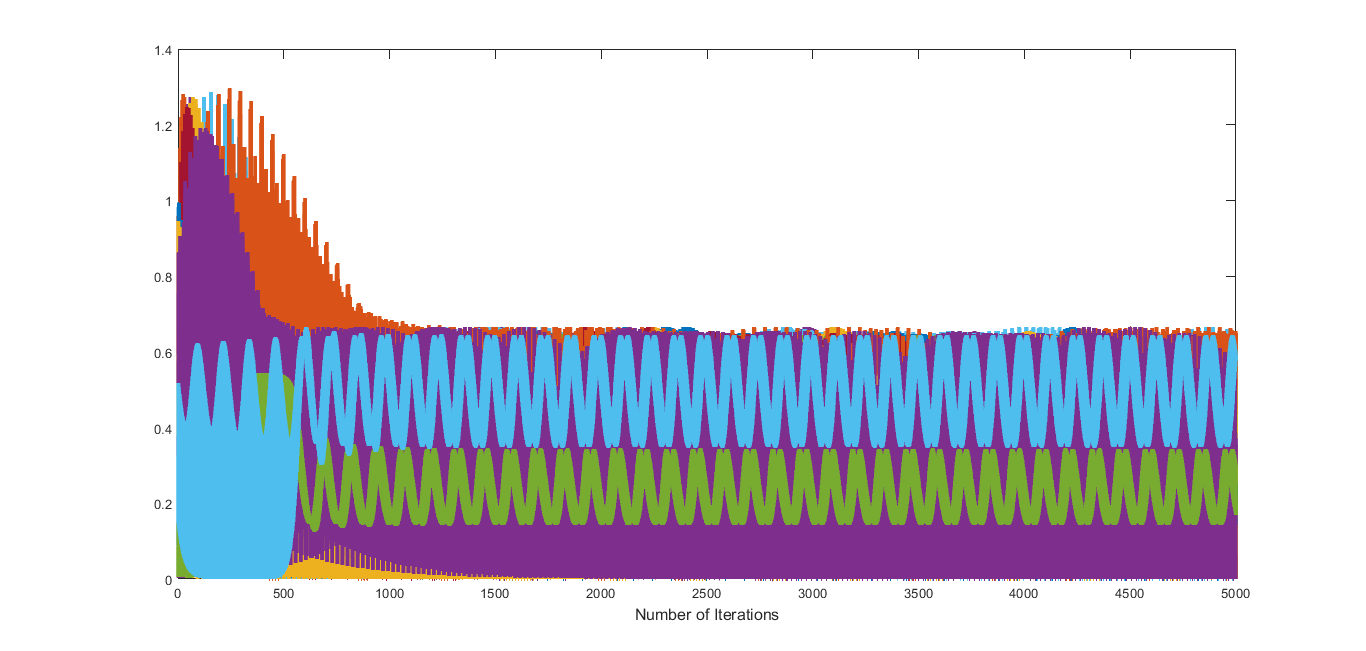}
\caption{Unstable (spiral source) phase plot of the equilibrium $P=(0.229, 0.528)$.}
\label{fig:fg10}
\end{center}
\end{figure}

\end{example}

\begin{example}
The parameters $r=0.0983, k=0.7337, a=0.4169, b=0.4608, c=0, d=0.5256$ and the delay f=25 satisfy \textit{Theorem-3}, consequently, the equilibrium $P=(0.7337, 0)$ is locally asymptotically stable as shown in Fig.\ref{fig:fg11}.

\begin{figure}[H]
\begin{center}
\includegraphics[scale=0.6]{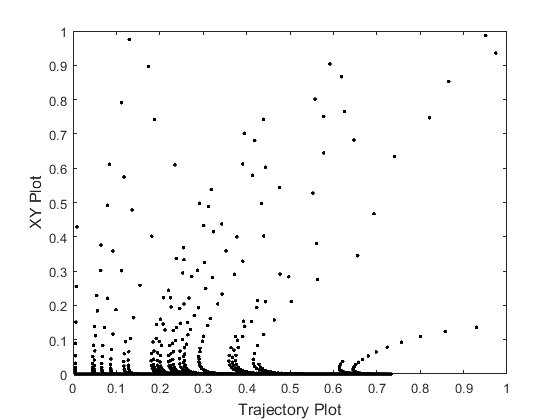}
\includegraphics[scale=0.6]{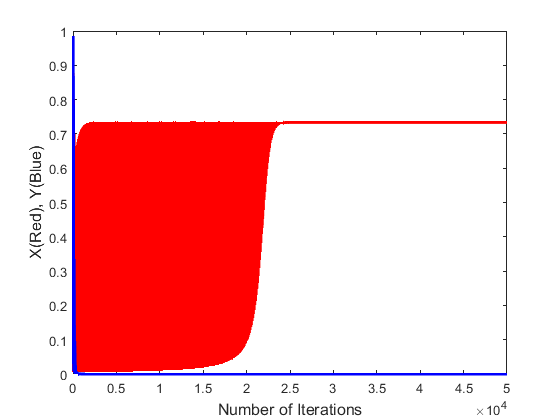}
\caption{Locally asymptotically stable phase and trajectory plot of the equilibrium $P=(0.7337, 0)$.}
\label{fig:fg11}
\end{center}
\end{figure}

\end{example}

\subsection{Fractality: Without Alee effect}
When $a$ and $c$ both are zero, then the system become classical, without any Alee effect. But role of the delay remains visible and that has been revealed using Fractal dimension of the trajectories. 

Here, we keep fixed the parameters of the system and vary the delay and initial values, then several trajectories were generated. Fractal dimension is employed to calculate the degree of fractality embedded in the prey and predator population trajectories (Supplementary File-1). 

\begin{figure}[H]
\begin{center}
\includegraphics[scale=0.515]{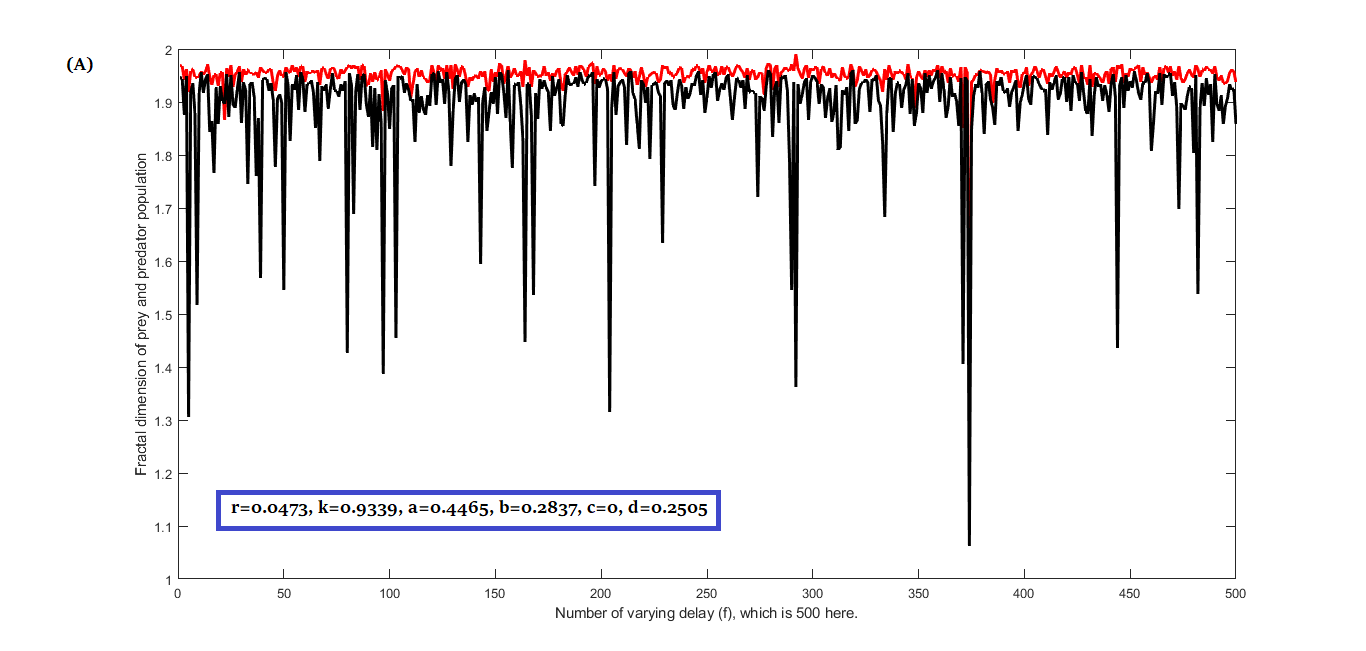}
\includegraphics[scale=0.515]{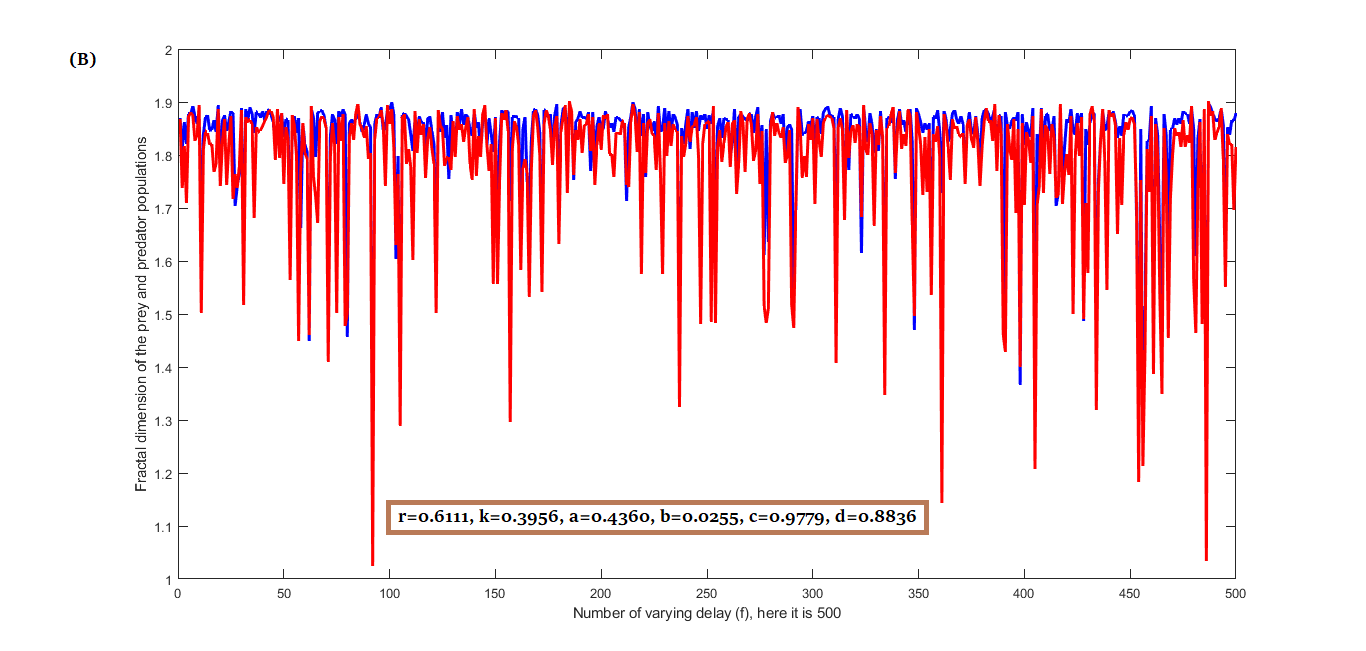}
\includegraphics[scale=0.515]{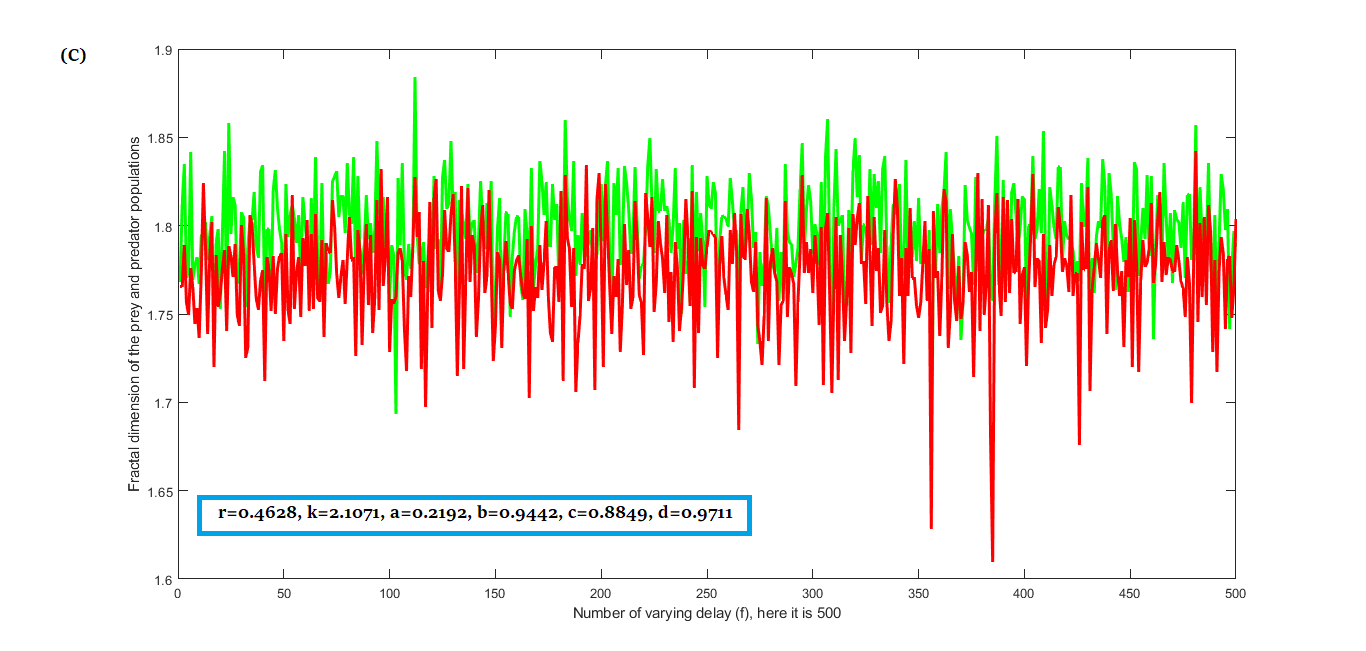}
\caption{Fractal dimension of the prey and predator population for 500 different delay parameters.}
\label{fig:fg12}
\end{center}
\end{figure}

\begin{figure}[H]
\begin{center}
\includegraphics[scale=0.515]{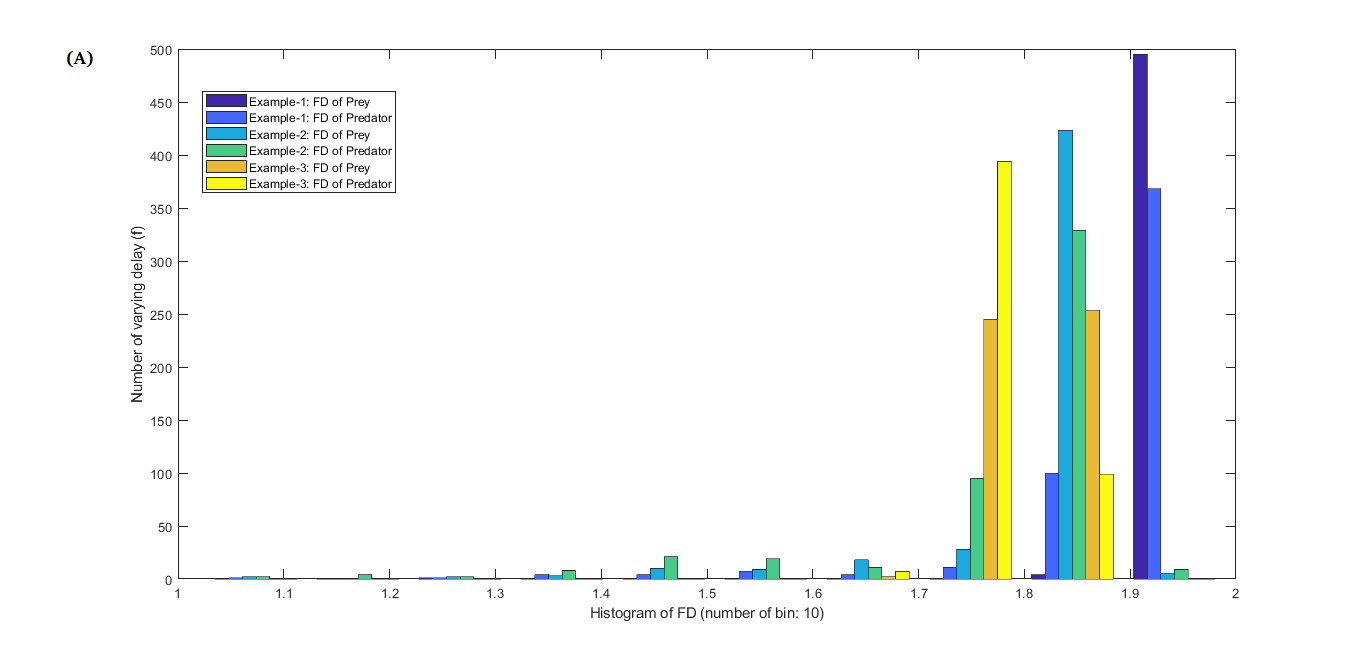}
\includegraphics[scale=0.515]{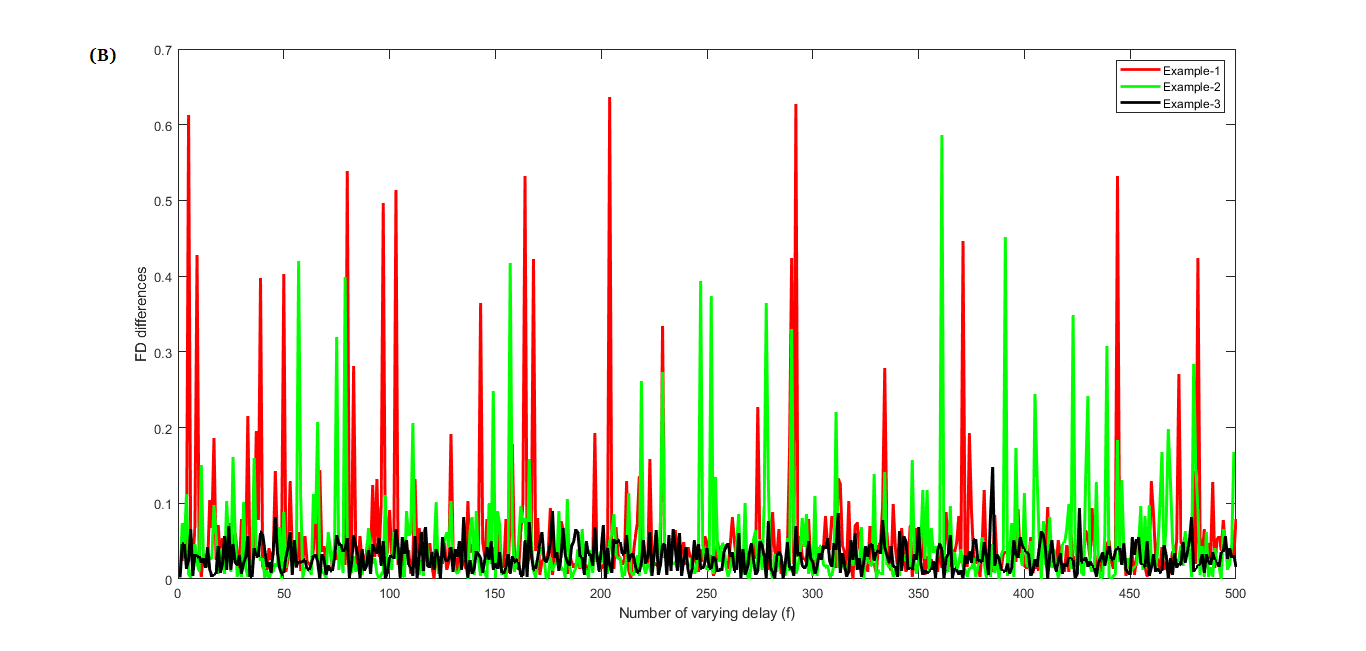}
\caption{(A): Difference between the FDs of prey and predator population for each 500 cases. (B): Histogram of FDs for three examples presented.}
\label{fig:fg13}
\end{center}
\end{figure}

\begin{itemize}
\item \textbf{Example-1:} Let the parameters $r=0.0473,$ $k=0.9339,$ $a=0.4465,$ $b=0.2837,$ $c=0$ and $d=0.2505$ be fixed for 500 different set of initial values and delay $(f)$. Trajectories in each case becomes unstable (highly-aperiodic), and fractal dimension for the prey and predator populations lie within the intervals [1.2548, 1.99] and [1.0627, 1.9605], respectively (Fig.\ref{fig:fg12}(A)). In most cases, FDs for prey as well as predator trajectories are cantering to within 1.92 and 1.95, signifying high degree of fractality (Fig.\ref{fig:fg13}(A)).     

\item \textbf{Example-2:} Let the parameters $r=0.6111,$ $k=0.3956,$ $a=0.4360,$ $b=0.0255,$ $c=0.9779$ and $d=0.8836$ be fixed for 500 different set of initial values and delay $(f)$. Trajectories in each case becomes unstable (quasi-periodic), and fractal dimension for the prey and predator populations lie within the intervals [1.0462, 1.8999] and [1.0255, 1.9002], respectively (Fig.\ref{fig:fg12}(B)). In most cases, FDs for prey as well as predator trajectories are cantering to within 1.835 and 1.86, implying moderate degree of fractality (Fig.\ref{fig:fg13}(A)).

\item \textbf{Example-3:} Let the parameters $r=0.4628,$ $k=2.1071,$ $a=0.2192,$ $b=0.9442,$ $c=0.8849$ and $d=0.9711$ be fixed for 500 different set of initial values and delay $(f)$. Trajectories in each case becomes unstable (quasi-periodic), and fractal dimension for the prey and predator populations lie within the intervals [1.6404, 1.8841] and [1.6096, 1.8422], respectively (Fig.\ref{fig:fg12}(C)). In most cases, FDs for prey as well as predator trajectories are cantering to within 1.775 and 1.7972, signifying high degree of fractality (Fig.\ref{fig:fg13}(A)).     
\end{itemize}

In the examples, it is observed that FDs of prey and predator trajectories are very near to each other, with small non-negative difference ((Fig.\ref{fig:fg13}(B)). Of note, even the difference between FDs also very disorderly in each example. Consequently, the trajectories of prey and predator populations with different delay (without any Alee effect) become unstable.   

Furthermore, it is noticed that FD of prey and predator population are getting minimum while delay (f) is significantly low. It turns that delay (f) is having a simple correlation with the FD of the prey and predator trajectories (Supplementary File-1).

\section{Discussions and Concluding Remarks}
The present study focuses on discrete dynamics of the delayed-prey-predator model with Alee effect. The system shows a wide variety of dynamics ranging from local asymptotic stability to quasi-periodic fractal trajectories. Alee effect on the dynamics also discussed and it was noticed in \textit{Example-10} that when predator Alee constant (c) becomes zero, then predator population asymptotically extinct. Furthermore, it is noticed that fractal dimensions of the prey and predator trajectories get close to 1 and 2, when the delay (f) is small and large enough, respectively. Also, fractal dimensions of prey and predator population trajectories with different delay show a cooperative characteristics of the prey and predator populations.

\bibliographystyle{elsarticle-num} 
\bibliography{cas-refs}

\section*{Appendix-1}

\begin{landscape}
$m_1=\frac{\left(8 a^2 b^3 k r^2-4 a b r (b k (b c-2)+d r-M) (b k (b c+r)+d r-M)+(-b k (b c+r-2)-d r+M) (b k (b c+r)+d r-M)^2\right) \exp \left(\frac{a \left(2 b^2 c k r (2 a b+b k+d)+b^4 c^2 k^2+r^2 (d-b k)^2-M^2\right)}{2 k (a b+d) \left(b r (2 a+k)+b^2 c k+d r-M\right)}\right)}{2 b k \left(b r (2 a+k)+b^2 c k+d r-M\right)^2}$\\\\

$$m_2=\frac{(-b k (b c+r)-d r+M) \exp \left(\frac{a \left(2 b^2 c k r (2 a b+b k+d)+b^4 c^2 k^2+r^2 (d-b k)^2-M^2\right)}{2 k (a b+d) \left(b r (2 a+k)+b^2 c k+d r-M\right)}\right)}{2 r}$$\\\\

$$m_3=\frac{\left(d (b k (r-b c)-d r+M)-2 a b^3 c k\right) \exp \left(\frac{2 b^2 c k r (2 a b+b k+d)+b^4 c^2 k^2+r^2 (d-b k)^2-M^2}{2 r (b k (b c+r)-d r+M)}\right)}{2 b k (a b+d)}$$\\\\

$m_4=\frac{\left(4 a^2 b^6 c^2 k^2+2 a b^3 c d k (b k (3 b c-r)+d r-M)+d \left(b^4 c^2 (2 d+1) k^2-2 b^3 c (d-1) k^2 r+b^2 k \left(2 c (d-1) (d r-M)+k r^2\right)+2 b k r (M-d r)+(M-d r)^2\right)\right) \exp \left(\frac{2 b^2 c k r (2 a b+b k+d)+b^4 c^2 k^2+r^2 (d-b k)^2-M^2}{2 r (b k (b c+r)-d r+M)}\right)}{d (b k (b c+r)-d r+M)^2}$	
\end{landscape}

\end{document}